\documentclass{svjour2}

\usepackage{graphicx}  
\usepackage{latexsym}
\usepackage{amssymb}

\newcommand{\beq}{\begin{equation}}
\newcommand{\eeq}{\end{equation}}
\newcommand{\sgn}{{\rm sgn}\,}
\newcommand{\bu}{\bar u}
\def\perpp{^{(\perp)}}
\def\parp{^{(\Vert )}}
\def\eqref#1{(\ref{#1})}

\journalname{General Relativity and Gravitation}

\begin{document}

\title{Position determination and strong field parallax effects for photon emitters in the Schwarzschild spacetime}

\author{Donato Bini \and 
Andrea Geralico \and Robert T. Jantzen
}

\institute{Donato Bini 
\at
Istituto per le Applicazioni del Calcolo ``M. Picone,'' CNR, Via dei Taurini 19, I-00185, Rome, Italy\\
ICRANet, Piazza della Repubblica 10, I-65122 Pescara, Italy\\
INFN, Sezione di Napoli,
Via Cintia, Edificio 6 - 80126 Napoli, Italy \\
\email{donato.bini@gmail.com} 
\and
Andrea Geralico 
\at
Istituto per le Applicazioni del Calcolo ``M. Picone,'' CNR, Via dei Taurini 19, I-00185, Rome, Italy\\
ICRANet, Piazza della Repubblica 10, I-65122 Pescara, Italy\\
\email{geralico@icra.it} 
\and
Robert T. Jantzen
\at   Department of Mathematics and Statistics, Villanova University, Villanova, PA 19085, USA   \\
ICRANet, Piazza della Repubblica 10, I-65122 Pescara, Italy\\
\email{robert.jantzen@villanova.edu}
}

\date{Received: date / Accepted: date / Version: \today}

\maketitle

\begin{abstract}
Position determination of photon emitters and associated 
strong field parallax effects are investigated
using relativistic optics when the photon orbits are confined to the equatorial  plane of the Schwarzschild spacetime. We assume the emitter is at a fixed space position and the receiver moves along a circular geodesic orbit. This study requires solving the inverse problem of determining the (spatial) intersection point of two null geodesic initial data problems, serving as a simplified model for applications in relativistic astrometry as well as in  radar and satellite communications. 
\end{abstract}

\section{Introduction}

The optics of photon motion in the Schwarzschild spacetime is well known and standardized in many textbooks in terms of an analytic representation using elliptic functions
(see e.g., 
Refs.~\cite{Darwin:1959,Misner:1974qy,Luminet:1979,Chandrasekhar:1985kt} and the more recent reviews 
\cite{Cadez:2005,Munoz:2014}),
but its application to parallax effects and stellar position determination has not received much attention. The present article aims to accomplish this goal in a fully relativistic strong field context, while simultaneously showing how these calculations
approach the more familiar ones of nearly flat spacetime in the weak field regime \cite{Arifov:1969,Semerak:2014kra}. 
Of course the approximation methods of space astrometry are sufficient to consider parallax effects of stars seen
from Earth or from space using the Gaia satellite project \cite{gaia} or future satellite measurements like LISA \cite{LISA}. However,
it is interesting in principle to investigate how one can handle them 
in strong gravitational fields with nontrivial curved spacetime optics.

To avoid excessive complication we
consider here the special case of photons emitted from a point in the same plane as an observer moving in a stable
circular orbit around a black hole or other compact object described by the Schwarzschild spacetime beyond some
minimal radius. Then relatively simple but curved 
2-dimensional spatial geometry can be used to model both the emitter, the 
receiver and photon orbits between them, a scenario which in turn can also be easily compared to the corresponding flat spacetime case.

We consider a stable timelike circular geodesic orbit $r=r_0\ge 6M$ in the equatorial plane $\theta=\pi/2$ of the Schwarzschild spacetime of mass $M$ receiving photons emitted from a point $P_*$ fixed in the spatial grid of the usual Boyer-Lindquist coordinates $(t,r,\theta,\phi)$ in that plane and study how to recover the emitter point coordinates $(r_*,\phi_*)$ from the angle of reception $\beta_0$ with respect to the azimuthal coordinate direction and the location $\phi_0$ on the receiving circle for two incoming such photons. The null geodesics in the Schwarzschild equatorial plane can be expressed giving the polar angle as a function of the radial coordinate using elliptic functions and parametrized by the conserved angular momentum along each photon trajectory. The latter in turn can be expressed in terms of the arrival angle $\beta_0$. One then has to (numerically) solve a pair of elliptic function relationships for the desired coordinates  $(r_*,\phi_*)$ given two sets of values for  $(\phi_0,\beta_0)$.

By using the orthonormal frame obtained by normalizing the coordinate frame vectors, corresponding to measurements by the static observers whose world lines are the time coordinate lines, stellar aberration effects due to the relative motion of the circularly orbiting observer are avoided, but can be calculated afterwards, see Appendix A. We start by setting up the geometry in the corresponding flat Minkowski spacetime using the same coordinate grid and will use this comparison later to consider the linearized corrections to the exact curved photon motion compared to the flat case.

\section{The flat spacetime scenario}

Let $(t,x,y,z)$ be inertial coordinates on flat Minkowski spacetime (signature $-+++$) with corresponding spherical spatial coordinates $(t,r,\theta,\phi)$, so that the line element reads  $ds^2 = -dt^2 + dr^2 + r^2(d\theta^2+\sin^2\theta \,d\phi^2)$
and let ${n}=\partial_t$ be the 4-velocity of the corresponding inertial observers. 
The spatial spherical orthonormal frame is given by
\beq
e_{\hat r}=\partial_r\,,\quad 
e_{\hat \theta}=\frac1r\,\partial_\theta \,,\quad 
e_{\hat \phi}=\frac{1}{r}\,\partial_\phi\,.
\eeq

Consider an accelerated (timelike) circular orbit in the equatorial plane $\theta=\pi/2 $ of flat Minkowski spacetime in spherical coordinates $(t,r,\theta,\phi)$
\beq
  t=t\,,\  
  r=r_0\,,\
  \theta=\pi/2\,,\
  \phi=\Omega\, t\,,\
\eeq
with $\Omega>0$.
An observer moving along this circular orbit has 4-velocity
\beq
  U = \gamma(U,n)({n}+\nu(U,{n}))\,,\quad
    \nu(U,{n})=\nu(U,{n})^{\hat\phi}e_{\hat\phi}\,,\quad
    \nu(U,{n})^{\hat\phi}=\Omega\, r_0\,.
\eeq
We will later choose $\Omega$ to be the coordinate frequency for the corresponding circular geodesic orbit in the Schwarzschild spacetime at this radius (given a fixed mass $M$).

\begin{figure}[h]
\begin{center}
\includegraphics[scale=1]{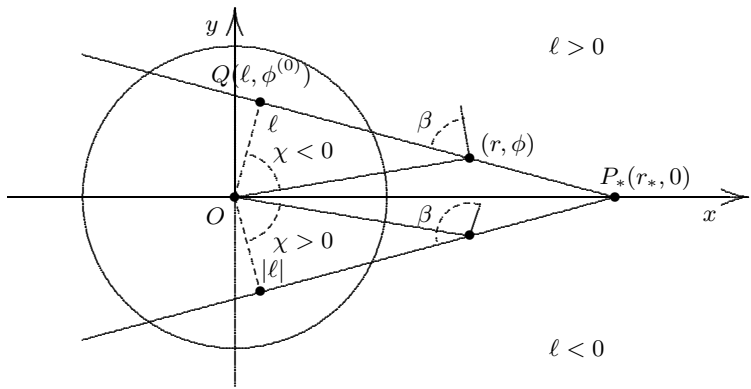}    
\end{center}
\vspace{1.5cm}
\caption{ The photon orbit in the $(r,\phi)$ plane from the point $P_*$ (chosen to have $\phi_*=0$) intersecting a circular orbit of radius $r_0$, shown for the case $\ell>0$ ($\ell<0$) above (below) the horizontal axis. 
The point $Q$ of closest approach to the origin has polar coordinates $(\ell, \phi^{(0)})$. For the part of the orbit from $P_*$ to $Q$, $r$ decreases to $\ell$ after which it increases. 
The counterclockwise angle $\beta$ of the incoming photon velocity with respect to the forward azimuthal direction is a negative acute angle decreasing to 0 when the photon orbit reaches $Q$ and then turns positive. 
The case $\ell<0$ corresponds to reflecting this diagram across the horizontal axis, so that the relative azimuthal direction is the  supplement of the corresponding previous case, and
$\beta$ is a negative obtuse angle decreasing to $-\pi$ at $Q$ and then taking values less than $-\pi$. 
One also sees that the difference angle $\chi=\phi-\phi^{(0)}=\beta<0$ for $\ell>0$ and $\chi=\pi+\beta>0$ for $\ell<0$.
\label{triangle0} }

\end{figure}

Equatorial plane null geodesics representing photons emitted from a point $P_*$, 
chosen here to lie on the 
positive $x$ axis only for convenience in our diagrams, arrive at each point of the circular orbit via a straight line trajectory, shown in Fig.~\ref{triangle0}. 
The geodesic equations describing these orbits parametrized by an affine parameter $\lambda$
(so that $P^\alpha=d x^\alpha/d\lambda$), are given by
\begin{eqnarray}
\frac{dt}{d\lambda}&=&E\,, \nonumber\\
\frac{dr}{d\lambda}&=& \pm  E\sqrt{1-\frac{\ell^2}{r^2} }\,,\nonumber\\
\frac{d\phi}{d\lambda}&=&  \frac{E\,\ell}{r^2} \,,
\end{eqnarray}
where the $\pm$ sign distinguishes the radially outgoing and ingoing directions. 
$E=E(P,n)$ is the conserved energy and $L_z\equiv E \ell$ is the conserved angular momentum  allowing their 4-momentum to be expressed in the form
\begin{eqnarray}\label{Pflat}
P&=&E ({n} +\hat \nu(P,{n}))
= E\left( \partial_t \pm \sqrt{1-\frac{\ell^2}{r^2}}\partial_r 
+ \frac{\ell}{r^2}\partial_\phi \right) \,,
\end{eqnarray}
with unit spatial velocity
\beq
\hat \nu(P,{n}) = \pm \sqrt{1-\frac{\ell^2}{r^2}}\,e_{\hat r} + \frac{\ell}{r}\,e_{\hat \phi}
\equiv \sin\beta\,e_{\hat r} + \cos\beta\,e_{\hat \phi}\,,
\eeq
defining the angle $\beta$ it makes with the azimuthal direction $e_{\hat \phi}$ in the clockwise direction
as shown in Fig.~\ref{triangle0},
so that $\beta<0$ as $r$ decreases. 

The 4-momentum $P$ of Eq.~\eqref{Pflat} referred to the rotating observer on the circular orbit (instead of the inertial observers of the coordinate grid) can be expressed  as
\begin{eqnarray}\label{PflatU}
P&=&E(P,U) (U +\hat \nu(P,U))\,,
\end{eqnarray}
with unit spatial velocity $ \hat\nu(P,U)$ with respect to $U$.
The two energies are related by
\begin{eqnarray}\label{doppler}
 E(P,U)&=& \gamma(U,n) E(P,n) \left(1-\nu(U,{n})\cdot \hat\nu(P,{n})\right)
\nonumber\\
  &=& \gamma(U,n) E (1-\Omega\, r_0\cos\beta)\,,
\end{eqnarray}
which describes the Doppler effect on the photon frequency if we set $E=\hbar \omega$, $E(P,U)=\hbar \omega(P,U)$. 
Eqs.~\eqref{Pflat}, \eqref{PflatU} may also be used to evaluate the angle $\beta_U$ of the incoming photon seen by the rotating observer due to aberration, 
a completely special relativistic effect discussed in detail in Appendix A.

Using the coordinate time  $t=E\lambda$ as a new parameter along the photon orbits, the geodesic equations reduce to
\begin{eqnarray}
\frac{dr}{dt}&=&  \pm \sqrt{1-\frac{\ell^2}{r^2}}\,,\qquad   \qquad
\frac{d\phi}{d t}=  \frac{\ell}{r^2}\,,
\end{eqnarray}
or introducing the reciprocal radius $u=1/r$ to compare with the Schwarzschild case below
\beq
\label{phot_orbit_phiu}
\frac{dt}{du}
=\mp \frac{1}{u^2\sqrt{1-\ell^2 u^2}}\,,\quad
\frac{d\phi}{du}
=\mp \frac{\ell}{\sqrt{1-\ell^2 u^2}}
=\mp  \frac{{\rm sgn}(\ell) }{\sqrt{\ell^{-2}-u^2}}
\,.
\eeq

\begin{figure}[h]
\begin{center}
\includegraphics[scale=1]{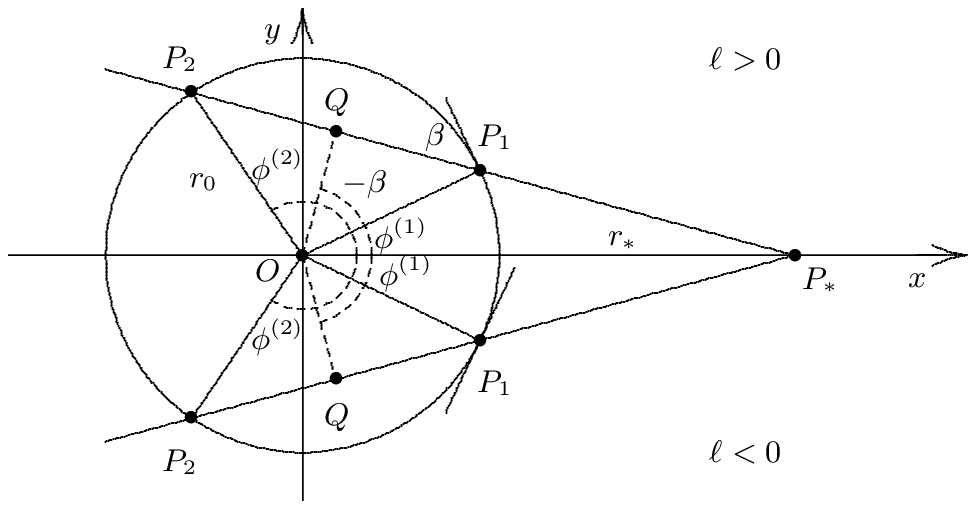}   
\end{center}
\vspace{1.5cm}
\caption{ The straight line photon orbit in the $(r,\phi)$ plane as in Fig.~1 connecting the point $P_*$ to $P_1$ and then $P_2$, shown with the angle of reception $\beta$ at $P_1$ on the circular orbit for $\ell>0$, and its complement for $\ell<0$.
The radial motion negative sign is appropriate for the segment $P_\ast Q$ and the plus sign for $Q P_2 $.
\label{triangle1} }

\end{figure}

The turning point $Q$ for the radial motion where this vanishes occurs at $r=|\ell|$, $\phi=\phi^{(0)}$, giving $|\ell|$ the interpretation as the impact parameter for this orbit (minimal radius), and requiring $r\ge |\ell|$. The time parameter measures the proper arclength along the photon trajectories, for which it is natural to choose the initial conditions at $t=t_0$ at the minimal radius
$r|_{t=t_0}=|\ell|, \phi|_{t=t_0}=\phi^{(0)}$, splitting the solutions into two segments, along which the radius is either monotonically increasing or decreasing with time, and is an even function of the difference $t-t_0$ or $\phi-\phi^{(0)}$.
The sign of $\ell$ introduces another sign factor into the rate of change of radius with respect to angle in Eq.~\eqref{phot_orbit_phiu}. 
Let $\phi^{(1)}$ denote the angle where the photon orbit first intersects the circular orbit (at $P_1$) and $\phi^{(2)}$ the second such intersection point ($P_2$), symmetrically located with respect to $Q$ and related by $\phi^{(0)}=(\phi^{(1)}+\phi^{(2)})/2$. Note that $\phi^{(0)}$, $\phi^{(1)}$, $\phi^{(2)}$ all depend on $\ell$, even if not explicitly shown in the notation.
Figs.~\ref{triangle0},\ref{triangle1} illustrate this situation.

From the geometry of Fig.~\ref{triangle0}, one can write down an arclength parametri\-zed straight line solution to the photon orbit equations in the polar coordinates.
If the photon orbit arrives at point $Q$ at $t=t_0$, since increments of $t$ measure arclength along the orbit, it is clear that the hypotenuse of the right triangle has length $|t-t_0|$, so 
\beq\label{rphi-t}
  r(t)=\sqrt{\ell^2+(t-t_0)^2} \,,\quad
	 \tan(\phi-\phi^{(0)}) = \frac{t-t_0}{\ell} 
	\,.
\eeq

Similarly, referring to the same Fig.~\ref{triangle0}, one can introduce a new angle $\chi=\phi-\phi^{(0)}$ along the photon orbit for which $\cos\chi=|\ell|/r=|\ell| u$ and so
\beq
  r(\phi) = \frac{|\ell|}{\cos\chi}
 = \frac{|\ell|}{\cos(\phi-\phi^{(0)})}\,,
\eeq 
or equivalently
\beq
 |\phi-\phi^{(0)}| =\arccos(|\ell| u)\,.
\eeq
As follows from the geometry of Figs.~\ref{triangle0} and \ref{triangle1},
for $\ell>0$, $\chi<0$ increases along the photon orbit from 
$\chi_*=\phi_*-\phi^{(0)}<0$ to 0 at $\phi=\phi^{(0)}$ (decreasing $r$) and then goes positive (increasing $r$).
For $\ell<0$, $\chi$ decreases from $\chi_*>0$ to 0 at $\phi=\phi^{(0)}$ (decreasing $r$) and then goes negative (increasing $r$).
In  both cases $|\chi_*|<\pi/2$.

On the other hand the orbital equations \eqref{phot_orbit_phiu} directly suggest the change of variable $\cos\chi=|\ell|u$,
 so that  $-\sin\chi\, d\chi = |\ell| du$, 
and then along the photon orbit the angular equation
for $\ell>0$ reduces  to
${d\phi}/{d\chi}=1$
as it should,
while the time equation integrates to
\beq
t=t_0+\ell\tan\chi\,,
\eeq
and hence
\beq
t=t_0\pm\frac{1}{u}\sqrt{1-\ell^2 u^2}\,,
\eeq
reproducing the radial solution \eqref{rphi-t}.

To compare the present simple, flat spacetime discussion with the more complicated change of variable needed for the Schwarzschild case, we consider the roots 
 $u_1=-{1}/{|\ell |}<0$ and $u_2={1}/{|\ell |}$  of the key quadratic expression appearing in the orbital equations
\beq
1-u^2 \ell^2=- \ell^2 (u-u_1) (u-u_2)\,,
\eeq
which leads to the following change of variable using the new angle $\bar\chi$ supplementary to $\chi$, namely
$\cos(\pi-\bar\chi)=-\cos\bar\chi= |\ell|/r= |\ell| u =\cos\chi$ which is equivalent to
\beq
\sin\left(\frac{ \bar\chi}2\right)= \sqrt{\frac{\bu-\bu_1}{\bu_2-\bu_1}}
\,,
\eeq
or
\begin{eqnarray}\label{u-chi-flat}
u&=&u_1+(u_2-u_1)\sin^2 \frac{\bar\chi}{2}
=-\frac{1}{|\ell |}+\frac{2}{|\ell |}\sin^2 \frac{\bar\chi}{2}
=-\frac{\cos \bar\chi}{|\ell |} \,,\nonumber\\    
du &=& \frac{\sin \bar\chi}{|\ell |} d\bar\chi\,.
\end{eqnarray}
This is the key to integrating the corresponding Schwarzschild equation.

For $\ell>0$, the geodesic formula can be rewritten in terms of the data of reception at the point $P_1$ where $\phi^{(0)}=\phi^{(1)}+\beta^{(1)}$ and $\ell=r_0 \cos\beta^{(1)}$,
namely
\beq
  r(\phi) = \frac{r_0 \cos\beta^{(1)}}{\cos(\beta^{(1)}+\phi^{(1)}-\phi)} \,.
\eeq
In order to compare the situation with multiple light ray signals received along the circular orbit,
we can evaluate this at the point $P_*(r_*,\phi_*)$ for two distinct photon trajectories, one has
\beq
  r_* = \frac{r_0 \cos\beta^{(1)}_1}{\cos(\beta^{(1)}_1+\phi^{(1)}_1-\phi_*)} 
      = \frac{r_0 \cos\beta^{(1)}_2}{\cos(\beta^{(1)}_2+\phi^{(1)}_2-\phi_*)} 
\,,
\eeq
where for convenience we had set $\phi_*=0$ before reaching this point.
For $\ell>0$, one must replace $\beta^{(1)}$ by $\pi-\beta^{(1)}$ and take the absolute value in the numerator.
These two equations can be used to first isolate $\phi_*$ and then back substitute into either one to get $r_*$.
One finds
\beq
   \tan\phi_* = -\frac{\cos\beta_1 \cos(\beta_2+\phi^{(1)}_2) -\cos\beta_2 \cos(\beta_1+\phi^{(1)}_1) }
                      {\cos\beta_1 \sin(\beta_2+\phi^{(1)}_2) -\cos\beta_2 \sin(\beta_1+\phi^{(1)}_1) }
\,.
\eeq

\begin{figure}[h]
\begin{center}
\includegraphics[scale=1]{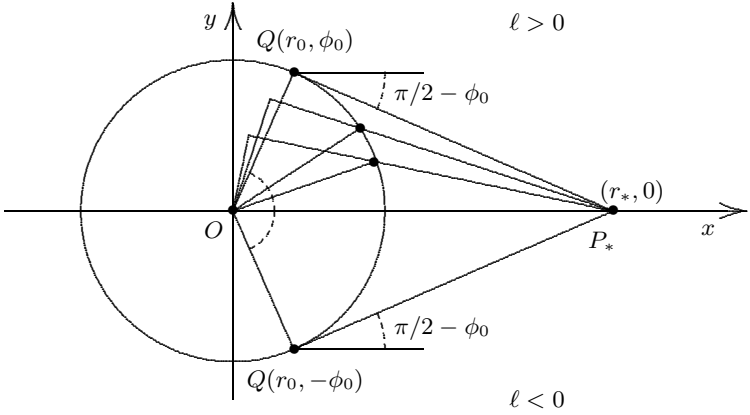}    
\end{center}
\vspace{1.5cm}
\caption{ 
The two photon orbits intersecting the circular orbit at the points $Q$ with maximum impact parameter $r_0=|\ell|$ and $\phi^{(0)}=\pm \phi_0$
correspond to $\chi=\phi^{(0)}-\phi^{(1)}=0$ and $\beta=0$ ($\ell>0$) and $\beta=-\pi$ ($\ell<0$) at the circular orbit, 
while zero impact parameter $\ell=0$ corresponds to $\phi^{(0)}=0$. 
Note that $r_*=r_0/\cos\phi_0$.
The parallax angle $\pi/2-\phi_0$ of the maximum impact parameter photons about the horizontal direction corresponding to photons arriving horizontally from infinite distance is shown, together with  
two intermediate $\ell>0$ photon orbits up to their point of closest approach to the origin.
\label{triangle2} }

\end{figure}

Alternatively this equality can be written
\beq
   r_* = \frac{\ell_1}{\cos(\phi^{(0)}(\ell_1)-\phi_*)} 
       = \frac{\ell_2}{\cos(\phi^{(0)}(\ell_2)-\phi_*)} \,.
\eeq
By 
1) isolating $\phi^{(0)}(\ell_1)-\phi_*,\phi^{(0)}(\ell_2)-\phi_*$, 
2) taking the cosine of their difference and expanding, 
3) eliminating the square roots by appropriately squaring, 
and 4) square rooting, one finds
\beq
  r_* = \frac{\sqrt{\ell_1^2+\ell_2^2-2\ell_1\ell_2 \cos\delta_{12}}}{|\sin\delta_{12}|} \,,
\eeq
where
\beq
\delta_{12}=\phi^{(0)}(\ell_2)-\phi^{(0)}(\ell_1)\,.
\eeq
Recalling that $\ell_1=r_0\cos\beta_1$, $\ell_1=r_0\cos\beta_2$, one obtains 
\beq
  r_* = \frac{\sqrt{\cos^2\beta_1+\cos^2\beta_2-2\cos\beta_1\cos\beta_2 \cos\delta_{12}}}
{|\sin\delta_{12}|} \,.
\eeq
This final result enables one to find the distance in terms of the measured azimuthal angles and photon arrival angles at two points of the orbit where those photons (both with $\ell>0$ for simplicity) first encounter the observer circular orbit. One can easily extend this to the case with any value of $\ell$.

\section{The Schwarzschild case}

Now consider a timelike circular geodesic orbit in the equatorial plane $\theta=\pi/2 $ of a Schwarzschild spacetime of mass $M$ in Boyer-Lindquist coordinates $(t,r,\theta,\phi)$
with metric
\beq
ds^2 = -(1-2M/r)\,dt^2 + (1-2M/r)^{-1} dr^2 + r^2(d\theta^2+\sin^2\theta \,d\phi^2)\,,
\eeq
namely
\beq
  t=t\,,\  
  r=r_0\,,\
  \theta=\pi/2\,,\
  \phi=\Omega\, t\,,\
\eeq
with $\Omega=\sqrt{M/r_0^3}$
and introduce the family of static observers along this orbit with the corresponding orthonormal frame defined everywhere by
\beq
{n}=\frac1{\sqrt{1-2M/r}}\,\partial_t \,,\quad 
e_{\hat r}= {\sqrt{1-2M/r}}\,\partial_r\,,\quad 
e_{\hat \theta}=\frac1r \,\partial_\theta \,,\quad 
e_{\hat \phi}=\frac{1}{r}\,\partial_\phi\,.
\eeq
An observer along the circular orbit has 4-velocity
\beq
  U = \gamma({n}+\nu(U,{n}))\,,\quad
    \nu(U,{n})=\nu(U,{n})^{\hat\phi}e_{\hat\phi}\,,\quad
    \nu(U,{n})^{\hat\phi}=\Omega\, r_0\,.
\eeq
Let $(x,y,z)$ be the natural Cartesian coordinates defined in terms of the spherical coordinates in flat spacetime. Replacing $M$ by $\epsilon M$  with $\epsilon=1$ in all quantities derived from the metric allows one to the make a series expansion in $\epsilon$ about 0 to see corrections to flat spacetime optics at $\epsilon=0$ in the weak field limit $M/r\ll 1$.

Equatorial plane null geodesics representing photons emitted from a point $P_*$ on the 
positive $x$ axis, chosen again for the convenience of our diagrams, arrive at each point of the circular orbit.  
Their orbits have conserved energy $E$ and angular momentum $L_z\equiv E \ell$, leading to the geodesic equations parametrized by an affine parameter $\lambda$ 
\begin{eqnarray}
\frac{dt}{d\lambda}&=&\frac{E}{1-{2M}/{r}}\,, \nonumber\\
\frac{dr}{d\lambda}&=& \pm E\sqrt{1-\frac{\ell^2}{r^2} \left(1-\frac{2M}{r} \right)}\,,\nonumber\\
\frac{d\phi}{d\lambda}&=& \frac{E \ell}{r^2}\,.
\end{eqnarray}
Their 4-momentum can then be expressed with respect to ${n}$ in the form
\begin{eqnarray}
P&=& E(P,{n}) ({n} +\hat \nu(P,{n}))\,,\quad
E(P,{n})=\frac{E}{\sqrt{1-2M/r}}\,,
\end{eqnarray}
with unit spatial velocity
\beq
\hat \nu(P,{n}) = \pm \sqrt{1-\frac{\ell^2}{r^2}\left(1-\frac{2M}{r} \right) }\,e_{\hat r} + \frac{\ell}{r} \sqrt{1-\frac{2M}{r}} \,e_{\hat \phi}
\equiv \sin\beta\,e_{\hat r} + \cos\beta\,e_{\hat \phi}\,,
\eeq
defining the angle $\beta$ it makes with the azimuthal direction $e_{\hat \phi}$ in the counterclockwise direction
as shown in Fig.~\ref{triangle0} for the corresponding flat spacetime case. 
When $\cos\beta=1$, then $r$ equals the impact parameter radius $R$ related to $\ell$ by
\beq \label{ell-R}
   \ell= \frac{R}{\sqrt{1-2M/R}} \quad \hbox{or}\quad
   \bar\ell= \frac{\bar R}{\sqrt{1-2/\bar R}}\,, 
\eeq
where $\bar\ell=\ell/M$ and $\bar R=R/M$ are dimensionless parameters,
with inverse
\beq 
{\bar R} = {\sqrt{1+\bar\ell^2}+1} \,.
\eeq
At the point of reception on the circular orbit at $r=r_0$, this instead allows us to find the angular momentum of the photon which arrives at a given angle $\beta_0$
\beq\label{ellbeta}
  \cos\beta_0 =\frac{\ell}{r_0} \sqrt{1-\frac{2M}{r_0}}\,.
\eeq

Note that the intrinsic curvature of this circular orbit in the curved Schwarzschild spacetime can be described in terms of the associated Lie relative curvature and radius of curvature defined in \cite{idcf1,idcf2}
\beq
  k_{\rm(lie)}= -\frac1r \sqrt{1-\frac{2M}{r}} \,,\quad
	R_{\rm(lie)}= \frac{1}{|k_{\rm(lie)}|}\,,
\eeq
where $R_{\rm(lie)}$ represents an \lq\lq effective" curvature radius of the orbit.
This restores the form of the flat spacetime formulas
\beq
\hat \nu(P,{n}) = \pm \sqrt{1-\frac{\ell^2}{R_{\rm(lie)}^2} }\,e_{\hat r} + \frac{\ell}{R_{\rm(lie)}}  \,e_{\hat \phi}\,.
\eeq

The reparametrization $ \lambda \to E \lambda$ is equivalent to setting $E=1$, leading to
\begin{eqnarray}
\frac{dt}{d\lambda}&=&\frac{1}{1-{2M}/{r}}\,, \nonumber\\
\frac{dr}{d\lambda}&=& \pm \sqrt{1-\frac{\ell^2}{r^2} \left(1-\frac{2M}{r} \right) }
\,,\nonumber\\
\frac{d\phi}{d\lambda}&=& \frac{\ell}{r^2}\,.
\end{eqnarray}
These imply the orbit equation
\beq
\frac{dr}{d\phi}=  \pm \frac{r^2}{\ell}\sqrt{1-\frac{\ell^2}{r^2} \left(1-\frac{2M}{r} \right) }\,.
\eeq
Introducing $u=1/r$ as in flat spacetime and the corresponding dimensionless variable $\bu=M/r=1/\bar r$, then $u \ell =\bu \bar\ell$ and
\beq
\label{dphi_du}
\frac{d\phi}{d\bu}=  \mp \frac{\bar \ell}{\sqrt{1-\bar \ell^2 \bu^2   \left(1-2\epsilon \bu  \right)} }
=\mp  \frac{\sgn (\bar\ell)}{\sqrt{ {\mathcal V}_{(\bar \ell, \epsilon)}(\bar u) }}\,,
\eeq
where 
\beq
{\mathcal V}_{(\bar \ell, \epsilon)}(\bar u)={1}/{\bar \ell^2}-\bu^2+2\epsilon \bu ^3\,.
\eeq
Inserting the factor $\epsilon=1$ for this Schwarzschild case allows us to consider the 
weak field limit $\bu=M/r\ll 1$,
letting $\epsilon \to0$ to compare with the corresponding flat spacetime result. 
The geodesic equation for the time variable can be similarly written as
\beq
\label{dt_du}
\frac1M \frac{dt}{d\bar u}=\mp \frac{1}{\bar u^2 (1-2\bar u)\ell } \frac{\sgn (\bar\ell)}{\sqrt{{\mathcal V}_{(\bar \ell, \epsilon)}(\bar u)}}\,.
\eeq
Both equations \eqref{dphi_du} and \eqref{dt_du} can be integrated exactly in terms of elliptic functions as shown below.

\subsection{Noncapture case}

For simplicity in what follows, we consider the case $|\bar\ell|>3\sqrt{3}$ of photons which arrive at the circular orbit with $|\phi|<\pi/2$ corresponding to the point $P_1$ of Figs. 1 and 2, but which are not captured by the black hole, and thus continue out to spatial infinity through points $Q$ and $P_2$ in a symmetrical way.
This introduces a significant change in the flat space picture in comparison, requiring  the consideration of values $|\phi_0|$ above some minimal value where capture first occurs as one decreases the angular momentum. A separate analysis of the geodesic formulas is required for the capture case.

The cubic polynomial in the denominator ${\mathcal V}_{(\bar \ell, \epsilon)}(\bar u)$  of Eq.~\eqref{dphi_du}
\beq
\label{cubic}
{\mathcal V}_{(\bar \ell, \epsilon)}(\bar u)=\frac{1}{\bar \ell^2}-\bu ^2+2\epsilon \bu ^3
=2\epsilon (\bu-\bu_1)(\bu-\bu_2)(\bu-\bu_3)
\eeq
admits three real roots, namely
\begin{eqnarray}\label{roots}
\epsilon \bu_1&=& \frac16 \left(1-e^{-i\pi/3}X -e^{i\pi/3}\frac1{X} \right)\,,\nonumber\\
\epsilon \bu_2&=& \frac16 \left(1-e^{i\pi/3}X -e^{-i\pi/3}\frac1{X} \right)\,,\nonumber\\
\epsilon \bu_3&=& \frac16 \left(1+X+\frac1{X} \right)\,,
\end{eqnarray}
where
\beq
X=\frac{1}{ (\bar\ell/\epsilon)^{2/3}}\left[(\bar\ell/\epsilon)^2-54  +6 i \sqrt{3}\sqrt{(\bar\ell/\epsilon)^2-27  }  \right]^{1/3}
\eeq
and these are ordered by $ \bu_1\le \bu_2\le \bu_3$, which fixes the value of the complex cube root here.
Introducing the notation
\beq
\bar \ell/\epsilon =3 \sqrt{3} \cosh \sigma \ge 3 \sqrt{3}
\eeq
we find
\begin{eqnarray}\label{xdef}
X&=&
e^{i\pi/3} e^{2i \theta_X/3 }\,,
\end{eqnarray}
namely
\beq
\theta_X
=-\arctan(\sinh \sigma)
=-\arctan \left(\sqrt{\left(\frac{\bar \ell/\epsilon}{3 \sqrt{3}}\right)^2-1}  \right)\,,
\eeq
which fixes the multivaluedness of the cube root in $X$.
Note that when $\bar\ell=3 \sqrt{3}$, 
corresponding to the circular null geodesic orbit,
this leads to $\theta_X=0$ and hence $X=e^{i\pi/3}$.
In the general case we have then 
\begin{eqnarray}
\epsilon  \bu_1&=&  
\frac16 -\frac13 \cos \left(\frac{2 \theta_X-\pi}{3}  \right)\,,  \nonumber\\
\epsilon \bu_2&=&  
\frac16 -\frac13 \cos \left(\frac{2 \theta_X+2\pi}{3}  \right)\,, \nonumber\\
\epsilon \bu_3&=&  
\frac16+\frac13 \cos \left(\frac{2\theta_X+\pi}{3}  \right)  \,,
\end{eqnarray}
with
\beq
\bu_1+\bu_2+\bu_3=\frac1{2\epsilon} \,,\qquad 
\bu_1\bu_2\bu_3=-\frac{1}{2\epsilon\bar \ell^2}\,,\qquad  
\bu_1 \bu_3+\bu_1 \bu_2+\bu_2 \bu_3=0\,.
\eeq
In the limit of large $\bar\ell$ (or small $\epsilon$) we find
\begin{eqnarray}\label{u-roots}
\epsilon \bu_1&=&   -\frac{\epsilon}{\bar\ell}+\frac{\epsilon^2 }{\bar\ell^2}-\frac{5\epsilon^3}{2\bar\ell^3}+O\left(\frac{\epsilon^4}{\bar\ell^{4}}\right) \,,\nonumber\\ 
\epsilon \bu_2&=&  \frac{\epsilon}{\bar\ell}+\frac{\epsilon^2 }{\bar\ell^2}+\frac{5\epsilon^3}{2\bar\ell^3}+O\left(\frac{\epsilon^4}{\bar\ell^{4}}\right) \,,\nonumber\\
\epsilon \bu_3&=&\frac1{2}  -\frac{2\epsilon^2 }{\bar\ell^2} + O\left(\frac{\epsilon^4}{\bar\ell^{4}}\right)\,.
\end{eqnarray}
As $\epsilon\to0$, $\bu_3\to\infty$ leaving only two finite roots for which $\bu_1+\bu_2\to0$.
Note that for geodesics satisfying $\bar\ell \ge 3 \sqrt{3}$, $\bu$ is confined to the interval $0<\bu\le \bu_2$, with $\bu=\bu_2$ corresponding to the point of closest approach to the black hole.

\begin{figure}
\begin{center}
\includegraphics[scale=0.4]{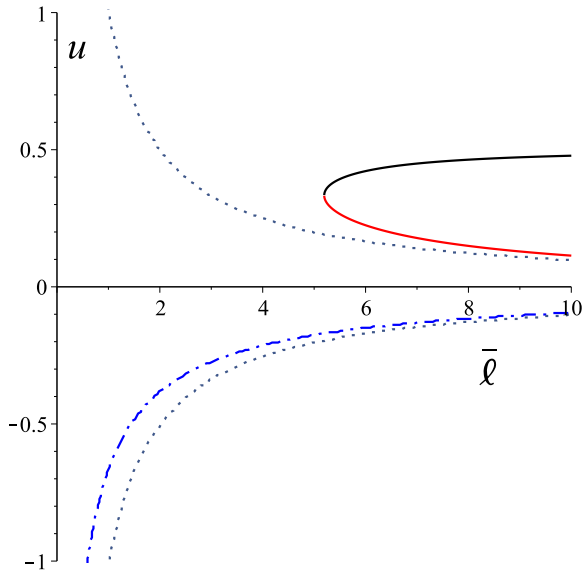}   
\end{center}
\caption{\label{fig:1} The solutions  of the cubic equation \eqref{cubic} for   $\bu$  are plotted as functions of $|\bar \ell|$ in the case $\epsilon=1$. For $|\bar \ell| > 3\sqrt{3}\approx 5.196$ 
the (positive) solutions are always two (solid curves)
and they coalesce (to $u=1/3$, corresponding to the circular photon orbit at $r=3M$) when $\bar \ell = 3\sqrt{3}$, but $\bu$ is confined to the interval $(0,\bu_2]$ along the trajectories, with the maximal value occurring at $\bu=\bu_2$ corresponding to the minimal value of the radius. The dotted curves correspond to the flat spacetime roots $\bu=\pm{1}/{|\bar \ell|}$ which are approached in the limit $\epsilon\to0$. }
\end{figure}

Integrating Eq.~\eqref{dphi_du} for the case $\bar\ell>0 $ to avoid sign complications leads to
\beq\label{eqphiu}
\phi(\bu)= \mp \sqrt{\frac{2}{\bu_3-\bu_1}}{\rm EllipticF}\left(\sin\left(\frac{ \chi}2\right), k  \right)+{\rm const.}
\eeq
with
\beq\label{sin-k}
\sin\left(\frac{ \chi}2\right)= \sqrt{\frac{\bu-\bu_1}{\bu_2-\bu_1}}\,,\qquad 
 k=\sqrt{\frac{\bu_2-\bu_1}{\bu_3-\bu_1}}\,,
\eeq
where EllipticF is the incomplete elliptic integral of the first kind defined explicitly in Appendix B.

To fix the additive constant in Eq.~\eqref{eqphiu}, consider the limit $\bu\to \bu_2$ where $\chi \to \pi$ and hence
\beq\label{eqphiu2}
\phi(\bu_2)= \mp \sqrt{\frac{2}{\bu_3-\bu_1}}{\rm EllipticK}(k)+{\rm const.}
\eeq
so that
\beq\label{phidiuExact}
\phi(\bu)-\phi(\bu_2)= \mp \sqrt{\frac{2}{\bu_3-\bu_1}} \left[{\rm EllipticF}\left(\sin\left(\frac{ \chi}2\right), k  \right)-{\rm EllipticK}(k)\right]\,,
\eeq
where the lower positive sign applies to the incoming photon orbit for $\bar\ell>0$, and EllipticK is the complete elliptic integral of the first kind  defined in Appendix B.
Hereafter we will only consider the incoming case with the positive sign.
This formula is directly analogous to the flat spacetime result \eqref{u-chi-flat} but where $\sin\chi/2$ is now related to $\phi$ by the more complicated composition with the elliptic function. In fact the $\epsilon\to0$ limit of \eqref{phidiuExact} with $k\to0$ shows how this limit comes about, using the values ${\rm EllipticF}(\sin\chi/2,0)=\chi/2$, ${\rm EllipticK}(0)=\pi/2$.
Note that this final result does not depend on our choice $\phi_* =0$ made for convenience of illustration, only that the photon orbit it describes has $\ell>0$.

Chandrasekhar \cite{Chandrasekhar:1985kt} (see pag. 131, Eq.~260) chooses $\phi(\bu_2)=0$, i.e., sets $\phi=0$ at the \lq\lq minimal distance." Here $\phi(\bu_2)$ corresponds to the flat spacetime impact parameter $\phi^{(0)}(\bar\ell)$, showing its explicit dependence on $\bar\ell$.
This elliptic function relationship can be inverted to give $u$ as a function of $\phi$ using the Jacobi elliptic function JacobiSN($x,k$) which inverts EllipticF($x,k$), first giving
\beq
\sin\frac{\chi}{2} = \sqrt{\frac{\bu-\bu_1}{\bu_2-\bu_1}}
={\rm JacobiSN}\left(\alpha(\phi),k\right) \,.
\eeq
with
\beq
\alpha(\phi)= \sqrt{\frac{\bu_3-\bu_1}{2}} \left( \phi-\phi(\bu_2)\right) +{\rm EllipticK}(k)\,,
\eeq
and then
\begin{eqnarray}\label{jacobi}
 \bu &=& \bu_1 + (\bu_2-\bu_1) {\rm JacobiSN}^2\left(\alpha(\phi),k\right)
\,.
\end{eqnarray}

\begin{figure}
\begin{center}
\includegraphics[scale=0.35]{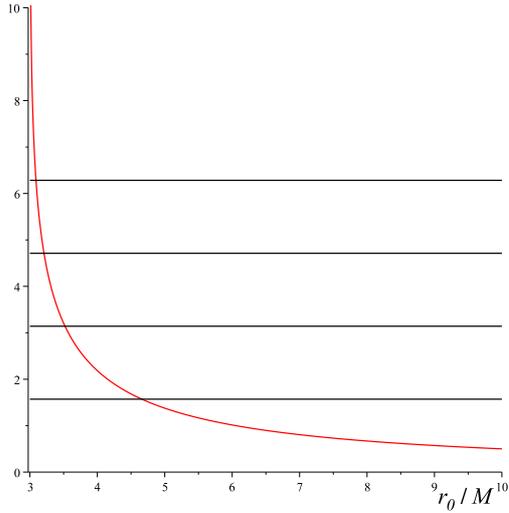} 
\end{center}  
\caption{\label{fig:4} 
The scattering angle for the photon orbit as a function of the minimal radius $r_0$ along the photon orbit. 
The horizontal lines mark off multiples of $\pi/2$.
For $\bar r_0=4.65$ this angle equals  $\pi/2$ corresponding to a 90 degree change in direction, while at smaller radii where it exceeds $\pi$ the photon makes at least one loop around the black hole horizon.  }
\end{figure}

Note that setting $u=0$ in the right hand side of \eqref{phidiuExact} for $\bar\ell>0$ and reversing its sign gives half the increase in $\phi$ over the entire photon orbit, so that by doubling this and subtracting $\pi$ gives the net scattering angle for the orbit.
Fig.~\eqref{fig:4} shows this scattering angle as a function of the minimal distance for $r/M\ge 3$. This approaches infinity as the photon orbits wrap around the circular photon orbit at $r/M=3$.

At the point of reception $(r_0=\bu_0^{-1},\phi_0)$ of a photon making an observed angle $\beta_0$ with the azimuthal direction, \eqref{ellbeta} gives the value of $\ell$ for the incoming photon orbit
\beq\label{barell-u}
  \bar\ell = \frac{\cos\beta_0}{\bu_0 \sqrt{1-2\bu_0}} \,.
\eeq
Then evaluating \eqref{phidiuExact} at $\bu=\bu_0$, $\phi(\bu_0)=\phi_0$
\beq\label{phidiuExact2}
\phi_0-\phi(\bu_2)=  
\sqrt{\frac{2}{\bu_3-\bu_1}} 
\left[{\rm EllipticF}\left(\sin\left(\frac{ \chi_0}2\right), k  \right)-{\rm EllipticK}(k)\right]\,,
\eeq
this determines the value $\phi(\bu_2)$.
Similarly at the point of emission $(\bar r_*=\bu_*^{-1},\phi_*)$, one has
\beq\label{phidiuExact3}
\phi_*-\phi(\bu_2)=   
\sqrt{\frac{2}{\bu_3-\bu_1}} 
\left[{\rm EllipticF}\left(\sin\left(\frac{ \chi_*}2\right), k  \right)-{\rm EllipticK}(k)\right]\,.
\eeq
For instance suppose we know that the point $P_*$ has $\phi_*=0$. Then Eq. \eqref{phidiuExact3} or its inverse Eq.~\eqref{jacobi} with $\phi_*=0$  determines $\bu_*$.

Suppose for the sake of an explicit example, we fix a relativistic Schwarz\-schild timelike circular geodesic orbit at $r_0=8M$ 
($\bu_0=1/8$) in the corresponding flat spacetime with initially $r_{*{\rm (flat)}}=20M$  and pick the corresponding value 
$\phi_0=\phi^{(0)}=\arccos(2/5)\approx66.4^\circ$, $\ell_{0{\rm (flat)}}=r_0$ (so that $\bar\ell_{0{\rm (flat)}} \bu_0=1$)
to be the last point $P_1=Q$ on the side of the orbit facing the emitter for $\phi>0$ (shown in Fig.~\ref{triangle2} for the flat case) which receives the incoming photon with angle $\beta_0=0$, such that
$1=\cos\beta_0=\ell_{0{\rm (flat)}}/r_0$ (and $\beta_{0*{\rm (flat)}}=-\phi_0$).

Next from
$1=\cos\beta_0=\bar\ell_0\bu_0\sqrt{1-2\bu_0} $  
calculate the corresponding Schwarz\-schild value 
$\bar\ell_0 =16/\sqrt{3}\approx9.24$ for that same incoming angle of $\beta_0=0$ for the curved orbit through this point $P_1$. Then using the null geodesic orbit equation \eqref{phidiuExact3} with $\phi(\bu_2)=\phi_0$ and $\phi_*=\phi(u_*)=0$,
calculate $r_*/M\approx 13.46=1/u_*$.  $\phi_*=0$ 
corresponds to an incoming (captured) photon with $\ell=0$
(frontal zero aberration point in the orbit).
Now one has the emitter point and the maximum negative acute angle 
$\beta_{0*}\approx-50.7^\circ$  for which photons will reach the circular orbit. 
Next calculate the minimum negative angle $\beta_{*\rm(cap)}\approx -69.1^\circ$
at which the photon is captured from
$\cos\beta_{*\rm(cap)}=3\sqrt{3}\bu_*\sqrt{1-2\bu_*} $.

Pick an intermediate value $ \beta_{*\rm(cap)} <\beta_{1*}<\beta_{0*}$  
for a new photon orbit  emitted from the point $P_*$,
say $\beta_{1*}=-65^\circ$ with angular momentum $\bar \ell_1\approx 6.16 $ determined by
$\cos\beta_{1*}= \bar\ell_1\bu_* \sqrt{1-2\bu_*} $.
Calculate the new angle $\phi_1\approx 17.1^\circ$ of reception and the corresponding incoming angle 
$-\pi/2<\beta_1\approx-55.8^\circ$.

Finally, to consider the time evolution of a given photon trajectory in spacetime, one has
 the explicit solution of Eq.~\eqref{dt_du} with the initial condition $t(\bu_*)=0$
\beq
t(\bar u) = H(\bar u) - H(\bar u_*)\,,
\eeq
where
\begin{eqnarray}
H(\bar u) 
&=& 
 -\frac{\bar\ell}{\bu} \sqrt{\mathcal V_{(\bar\ell,\epsilon)}(\bu)}
\nonumber\\
&& 
+C_1 {\rm EllipticPi}\left(\sin \frac{\chi}{2}, 1-\frac{\bu_2}{\bu_1},k\right)
\nonumber\\
&&
+C_2 {\rm EllipticPi}\left(\sin \frac{\chi}{2},  \frac{2(\bu_1-\bu_2)}{2\bu_1-1},k\right)
\nonumber\\
&&
+C_3 {\rm EllipticE}\left(\sin \frac{\chi}{2},k\right)
+C_4 {\rm EllipticF}\left(\sin \frac{\chi}{2},k\right)
\,,
\end{eqnarray}
with
\begin{eqnarray}
C_1 &=&-\frac{4}{\sqrt{2}\bar\ell\bu_1 \sqrt{\bu_3-\bu_1}}\,,
\quad
C_2 = \frac{4\sqrt{2}}{\bar \ell (2\bu_1-1)\sqrt{\bu_3-\bu_1}}\,,\nonumber\\
C_3 &=& \sqrt{2}\bar\ell(\bu_3-\bu_1)\,,
\quad
C_4 = \frac{1}{\sqrt{2}\bar \ell}\frac{1}{\bu_1 \bu_2\sqrt{\bu_3-\bu_1}}
\,,
\end{eqnarray}
recalling the definitions \eqref{sin-k} introduced above.

Imagining the curved orbit figure corresponding to the flat spacetime case illustrated Fig.~\ref{triangle2},
the horizontal direction depicted on the circular orbit at the maximal impact parameter cases still corresponds to photons arriving from spatial infinity.   
Compared to the celestial circle of the distant stars in the equatorial plane, the apparent direction of the point $P_*$ varies from $\beta=0$ at the lower point $Q$ in Fig.~\ref{triangle2} to $\beta=\pi$ at the upper point $Q$ as one moves along the circular orbit.

The parallax angle of $P_*$ is equal to the angle made by the incoming photon with the horizontal direction $\phi=0$ which corresponds to
$\beta=\pi/2-\phi_0$ at the lower point $Q$ for $\ell<0$ 
and
$\beta=\pi/2+\phi_0$ at the upper point $Q$ for $\ell>0$.  
However, nonrelativistic discussions define the parallax angle in terms of the photons observed when the point $P_2$ of the orbit lies at the extreme antipodal points
$\phi=\pm\pi/2$ on the circular orbit relative to the direction $\phi=0$. This requires a piecewise solution of the photon orbit to deal with the sign change after the minimal approach point $Q$ is reached inside the circular orbit. This leads to a more complicated discussion, which would describe the exact parallax for an emitter on the polar axis of the spherical coordinate system orthogonal to the equatorial plane.

\subsection{Capture case}

Orbits for $|\bar\ell|< 3\sqrt{3}\approx 5.20$ are captured by the black hole so no minimal radius exists, and one must therefore choose initial conditions differently from the noncapture case in \eqref{eqphiu2}.
The formulas \eqref{roots} remain valid but the root $u_1$ remains real and negative, while the roots $u_2$, $u_3$ become complex conjugates. In the definition of $X$ in \eqref{xdef} the quantity $\theta_X$ vanishes at $|\bar\ell|= 3\sqrt{3}$ and then becomes purely imaginary.  
We choose the initial condition $\phi(\bu_*)=0$ so that
\beq\label{phidiuExact4}
\phi(\bu)= 
 \sqrt{\frac{2}{\bu_3-\bu_1}} \left[{\rm EllipticF}\left(\sin\left(\frac{ \chi}2\right), k  \right)-  
{\rm EllipticF}\left(\sin\left(\frac{ \chi_*}2\right), k  \right)
\right]\,.
\eeq
For example, if $\bar\ell=5$ one finds 
$\bu_1 = -0.172$,
$\bu_2 = 0.336+0.0540 i=\bu_3^*$
and $\phi(\bu_0)\approx 16.3^\circ$.

\section{The inverse problem}

Given $r_*$, in order to determine the corresponding value $\phi_0=\phi(u_2)$ for which the point $Q$ is on the circular orbit,
one must first determine the value of $\ell>0$ from the condition \eqref{barell-u} with $\beta_0=0$ and then setting $\phi_*=0$
in \eqref{phidiuExact3} with this value of $\ell$, one finds the corresponding value of $\phi(u_2)$. In our above example, this leads to a parallax angle of $23.6^\circ$, the complement of the angle $\phi_0$ we started with.

Finally consider how to determine the radial position $r_*$ from two pairs of incoming data $(\phi,\beta)$  from the same point $P_*$,
namely $(\phi_1,\beta_1)$ and $(\phi_2,\beta_2)$ at $r=r_0$ which immediately gives us the corresponding values $\bar\ell_1$ and $\bar\ell_2$ which determine the roots $\bu_i(\bar\ell)$ describing the incoming photon.

Suppose we represent Eq.~\eqref{phidiuExact}  by
\beq
\phi(\bu)-\phi(\bu_2(\bar\ell))= F(\bu,\bar\ell)\,,
\eeq
where $F(\bu,\bar\ell)$ denotes symbolically the right-hand-side of that equation.
Evaluating it first at $\bu=\bu_0$ for $\bar\ell_1$ and $\bar\ell_2$ we then solve for 
$\phi(\bu_2(\bar\ell_1))$ and $\phi(\bu_2(\bar\ell_2))$. Then using these known values, we evaluate this equation again at 
$\bu=\bu_*$ for $\phi(\bu_*)=\phi_1$ for $\bar\ell_1$ and  $\phi(\bu_*)=\phi_2$ for $\bar\ell_2$.
These two independent equations for the two unknowns $\bu_*$, $\phi_*$ can be solved numerically.

We can use the data from our example as a test case to recover the coordinates of $P_*$. Starting from 
$(\phi_1,\beta_1)=(66.4^\circ,0^\circ)$ and $(\phi_2,\beta_2)=( 21.6^\circ,-48.1^\circ)$ following the above steps, one easily finds indeed $r_*=13.4 M$ and $\phi_*=0.00910^\circ$ keeping only 3 significant digits.

\section{Targeting problem}

Suppose both the emitter and receiver are in circular motion in the same plane. One can study the significantly more complicated problem of aiming an emitted photon in such a way as to hit the receiver. This requires including the time evolution of the photon.
As a first step one can track a photon emitted at $t=0$, $\phi_*=0$ from the point $P_*$ on a circular geodesic orbit at radius $r_*=M/u_*>r_0$ (with angular velocity $\Omega_{\rm(em)}=\sqrt{M/r_*^3}$)
and calculate the angle $\phi_0$ it arrives  at the receiver at time $t=t_1$ in the circular orbit $\phi=\Omega_{\rm(rec)} t$ at radius $r_0=M/u_0$ 
(with angular velocity $\Omega_{\rm(rec)}=\sqrt{M/r_0^3}$).

Let $\phi(u,\ell)$, $t(u,\ell)$ describe the null geodesic  from $P_*$ such that 
$\phi(u_*,\ell)=0$, $t(u_*,\ell)=0$.
The condition that the photon arrive at the same angle and time as the receiver is
\beq
 t_1\Omega_{\rm(rec)} = \phi(u_0,\ell)\,,\quad
 t_1 = t(u_0,\ell)\,.
\eeq
The ratio of these two relations then leads to
\beq
u_0^{3/2}=M\Omega_{\rm(rec)} = M \frac{\phi(u_0,\ell)}{ t(u_0,\ell)}\,.
\eeq
This can then be solved numerically for the value of $\ell$ required to fix the initial photon direction $\beta_*=\arccos(\bar\ell \bu_* \sqrt{1-2\bu_*})$.

Provided that the photon orbit can be described by the interval of $\ell>0$ values for a noncapture orbit, we can use the explicit elliptic function formulas derived above to determine the solution for $\ell$. In particular for the above noncapture example with $r_0=8M$, $r_*\approx13.46M$ one finds $\bar\ell\approx5.814$, $\beta_*\approx 66.5^\circ$, determining the photon orbit which arrives at the receiver at $\phi_0\approx 19.9^\circ$.

\section{Concluding remarks}

The investigation presented here shows that one can perform the strong field calculations of relativistic optics required to determine positions of photon emitters (like a star or satellite) from observations made by an observer in circular motion around a nonrotating compact object or black hole. Starting from the much simpler situation in flat spacetime optics for comparison, calculations are then performed for the corresponding Schwarzschild spacetime where both emitter and receiver are confined to the same plane. 
Given two observation points on the observer orbit, one can uniquely determine the location coordinates of an emitter fixed in the static grid of the spacetime.
This analysis also shows concretely how one can solve the inverse problem of using photons to target a particular object in relative motion.
By expanding exact formulas involving elliptic functions in the mass of the source of the gravitational field, one finds the corrections to the flat spacetime optics necessary to take into account spacetime curvature effects in the weak field case.
Generalizations of these calculations to the full 3-dimensional spatial geometry in the Schwarzschild spacetime as well as to the case of a rotating Kerr spacetime will be considered in future work.

\appendix

\section{The aberration map}

The (nonlinear) aberration map describes the relationship between the apparent
directions of a light ray as seen by two arbitrary observers in relative motion at the same spacetime point. These are purely special relativistic considerations in the tangent space valid both for any spacetime.
Consider a photon
with 4-momentum $P$ and unit relative velocities $\hat\nu(P,u) $ or $\hat\nu(P,U)$ 
\beq
\label{P_U_u}
    P = E(P,u) [ u + \hat\nu(P,u) ]  = E(P,U) [ U + \hat\nu(P,U) ]
\eeq
as seen by two different observers
with 4-velocities $u$ and 
\beq
 U=\gamma(U,u)  [ u + \nu(U,u) ]=\gamma(U,u)  [ u + ||\nu(U,u)||\hat \nu(U,u) ]\,,
\eeq
where in the last equation we have introduced the magnitude and the unit spatial direction of the relative velocity $\nu(U,u)=||\nu(U,u)||\hat \nu(U,u)$. 
Moreover, the inverse relations are easily written using the convenient but precise relative observer notation of Ref.~\cite{Jantzen:1992rg}, namely
\beq
u=\gamma(u,U) [U+\nu(u,U)]\,,
\eeq
with $\gamma(u,U)=\gamma(U,u)=-U\cdot u$ and 
\beq
-\nu(u,U)= \gamma (U,u) ||\nu(U,u)||\, \left[\, ||\nu(U,u)||\, u +\hat \nu(U,u)\right]\,.
\eeq
Contracting Eq.~\eqref{P_U_u} by $U$ one easily finds
\beq
    E(P,U) = \gamma(U,u) [ 1 - \nu(U,u) \cdot \hat\nu(P,u) ] E(P,u)\,,
\eeq
which can be rewritten in terms of the photon frequencies $\omega(P,u) = E(P,u) / \hbar$ and $\omega(P,U) = E(P,U) / \hbar$
leading to the formula \eqref{doppler}
for the relativistic Doppler effect. 

Instead projecting Eq.~\eqref{P_U_u} orthogonally to $u$ leads to
\beq
P(u,U)\hat\nu(P,U)= \frac{E(P,u)}{E(P,U)}\hat\nu(P,u)-\gamma(U,u) \nu(U,u)\,,
\eeq
where $P(u,U)= P(u) P(U)$ is the projection map from the local rest space of $U$ into that of $u$, and
$P(u)^\alpha{}_\beta =\delta^\alpha{}_\beta +u^\alpha u_\beta$. 
Expanding the left-hand-side of this equation, namely $P(u,U)\hat\nu(P,U)=P(u)\hat\nu(P,U)=\hat\nu(P,U)+u [u \cdot \hat\nu(P,U)]$, and 
re-expressing the scalar product $u \cdot \hat\nu(P,U)$ replacing $\hat\nu(P,U)$ by
\beq\label{nuPU}
\hat\nu(P,U)=\frac{E(P,u)}{E(P,U)}[u+\hat\nu(P,u)]-U 
\eeq
implies
\beq
\hat \nu(P,U) -\nu(u,U)= \frac{E(P,u)}{E(P,U)} P(u,U)^{-1}\hat \nu (P,u)
\eeq
or
\beq
\hat \nu (P,u)=\frac{E(P,U)}{E(P,u)} P(u,U) [\hat \nu(P,U) -\nu(u,U)]\,,
\eeq
with its inverse relation obtained by exchanging  $U$ and $u$,
\beq
\hat \nu (P,U)=\frac{E(P,u)}{E(P,U)} P(U,u) [\hat \nu(P,u) -\nu(U,u)]\,.
\eeq

More familiar relations  (for classical Doppler effect) are obtained decomposing the relative velocity $\hat\nu(P,U)$
into parts parallel and perpendicular to the direction $\hat\nu(u,U)$
of relative motion
\beq
\hat\nu(P,U) = \hat\nu(P,U)\parp + \hat\nu(P,U)\perpp\,,
\,   
\eeq 
with $\hat\nu(P,U)\parp=||\hat\nu(P,U)\parp||\hat\nu(u,U)$, and similarly
\beq
\hat\nu(P,u) = \hat\nu(P,u)\parp + \hat\nu(P,u)\perpp\,,
\,   
\eeq 
with $\hat\nu(P,u)\parp=||\hat\nu(P,u)\parp||\hat\nu(U,u)$.
Re-expressing the relative projection in terms of the boost 
leads to the formulas
\begin{eqnarray}
    \hat\nu(P,U)\perpp &=& \gamma(U,u)^{-1} [ 1 - \nu(U,u) \cdot \hat\nu(P,u) ]^{-1}
                         \hat\nu(P,u)\perpp\,, \nonumber\\
    \hat\nu(P,U)\parp &=& [ 1 - \nu(U,u) \cdot \hat\nu(P,u) ]^{-1}
                        B_{\rm (lrs)}(u,U) [ \hat\nu(P,u)\parp - \nu(U,u) ] \,,
\end{eqnarray}
where we have introduced the boost map $B_{\rm (lrs)}(u,U)$ 
from the local rest space of $U$ to that of $u$
(see Eq.~4.22 of Ref.~\cite{Jantzen:1992rg}) defined so that
\beq
[B_{\rm (lrs)}(u,U) S]^\alpha= \left[P(u)-\frac{\gamma(U,u)}{ 1+\gamma(U,u) }\nu(U,u)\otimes \nu(U,u)\right]{}^{\alpha}{}_\beta [P(u,U)S]^\beta
\eeq
for any vector $S$ belonging to the local rest space of $U$.

For the special case
\beq
             \hat \nu(P,u)\parp = 0 \ ,
           \qquad  || \hat \nu(P,u)\perpp || = 1
\eeq
of a light ray orthogonal to the relative motion as seen by $u$, one has
the familiar result
\beq
       \tan \theta_U = {|| \hat \nu(P,U)\parp || \over || \hat \nu(P,U)\perpp || }
      = \gamma(U,u)  ||\nu(U,u) ||
\eeq
for the angle away from the perpendicular direction as seen by $U$, using
the fact that boosts preserve lengths.

For the present application to the circularly rotating observer $U$ and the static observer $n$, let
\beq
 U = \gamma(U,n) [n + ||\nu(U,n) || e_{\hat\phi}]\,,\quad
 E_{\hat\phi} = \gamma(U,n) [||\nu(U,n) || n + e_{\hat\phi}]\,.
\eeq
The arriving photon can be viewed by either observer, defining the azimuthal angle of reception by
\beq
  \hat\nu(P,U)^{\hat\phi} = \hat\nu(P,U) \cdot E_{\hat\phi} = \cos\beta_U\,,\quad 
  \hat\nu(P,n)^{\hat\phi} = \hat\nu(P,n) \cdot e_{\hat\phi} = \cos\beta_n\,.
\eeq 
From Eq.~\eqref{nuPU} with $u=n$, contracting both sides with $ E_{\hat\phi}$, one finds 
\beq
  \cos\beta_U = \frac{\cos\beta_n-||\nu(U,n)||}{1-\cos\beta_n ||\nu(U,n)||}\,,
\eeq
where 
\beq
||\nu(U,n)|| = \frac{r_0 \Omega}{\sqrt{1-2M/r_0}} \equiv \Omega R_{\rm(lie)}^0\,.
\eeq
This is the aberration map at the reception point whose inverse is needed to transform the observed angle to the static  observer value required for the computations.

\section{Approximate formulas}

Here we summarize the linear expansion in $\epsilon$ of this analysis, noting again that setting $\epsilon=1$ in the resulting expressions is equivalent to the weak field limit $M/r\ll1$ of the exact expressions. This describes the corrections due to the curvature of spacetime.

We use the following definitions \cite{Gradshteyn} of the complete and incomplete elliptic integrals of the first kind
 ${\rm EllipticK}(k)$ and ${\rm EllipticF}(\varphi,k)$   defined respectively by
\begin{eqnarray}
{\rm EllipticF}(\varphi,k)&=&\int_0^{\varphi}\frac{dz}{\sqrt{1-k^2\sin^2z}}\,,\quad
{\rm EllipticK}(k)={\rm EllipticF}(\pi/2,k)
\,.
\end{eqnarray}

The following expansions of elliptic functions and their parameters are useful.
\begin{eqnarray}
\label{ellF_exp}
&&{\rm EllipticF}\left(x_0+\epsilon x_1, k_{1/2}\epsilon^{1/2}+k_{3/2}\epsilon^{3/2}  \right)
\nonumber\\ &&\qquad
= \arcsin x_0 +\frac{\epsilon}{4}
\left[ k_{1/2}^2 \arcsin x_0 +\frac{x_1}{\sqrt{1-x_0^2}}-x_0 k_{1/2}^2 \sqrt{1-x_0^2}   \right]+O(\epsilon^2) \,,
\nonumber\\
&& {\rm EllipticK}(k_{1/2}\epsilon^{1/2}+k_{3/2}\epsilon^{3/2})=\frac{\pi}{2}\left(1+\frac14 k_{1/2}^2\epsilon \right)+O(\epsilon^2)
\,.
\end{eqnarray}

From the definition \eqref{sin-k}, it follows from \eqref{u-roots} that the parameter $k\to0$ as $\epsilon\to0$ and
$k^2=1$ when $\bar\ell=3\sqrt{3} \epsilon$ while, for large $\bar\ell$
\beq
k^2=\frac{4\epsilon}{\bar\ell}-\frac{8\epsilon^2}{\bar\ell^2}+\frac{50\epsilon^3}{\bar\ell^3}+O\left(\frac{\epsilon^4}{\bar\ell^{4}}\right)\,.
\eeq
Moreover, approximate expressions for $\sin \chi/2$ and ${\rm EllipticK}(k)$ are the following
\begin{eqnarray}
&&\sin\frac\chi{2} = \sqrt{\frac{\bar\ell \bu }{2\epsilon}}\left(1+\frac{1}{2\bu }\frac{\epsilon}{\bar\ell} -\frac{1+4\bu +10\bu ^2}{8\bu ^2} \frac{\epsilon^2}{\bar\ell^2} +\frac{1+4\bu +10\bu ^2}{16\bu ^3} \frac{\epsilon^3}{\bar\ell^3}  \right) +O\left( \epsilon^{7/2}  \right)
\,,\nonumber\\
&&{\rm EllipticK}(k) =\frac{\pi}{2}\left(1+\frac{\epsilon}{\bar\ell}+\frac14  \frac{\epsilon^2}{\bar\ell^2}\right)+O(\epsilon^3)\,.
\end{eqnarray}

Using the approximate formula \eqref{ellF_exp} in Eq.~\eqref{eqphiu2} with the plus sign for the incoming photon,
we find for $\bar\ell>0$
\beq
\phi(\bu)-\phi(\bu_2) =  
\left(2\arcsin\left(\sqrt{\frac{1+\bu\bar\ell}{2}} \right) -\frac{\epsilon}{\bar\ell}\left[ \sqrt{1-\bu^2 \bar\ell^2}+\frac{1}{\sqrt{1-\bu^2 \bar\ell^2}}\right]\right)+O(\epsilon^2)
\eeq
or using the identity
\beq
\arccos x= \pi -2\arcsin \left( \sqrt{\frac{1+x}{2}} \right)\,,
\eeq
we get
\beq
\phi(\bu)-\phi(\bu_2)= \Phi_{(\bar\ell,\epsilon)}(\bu) +O(\epsilon^2)\,,
\eeq
where
\beq\label{Phiepsilon}
\Phi_{(\bar\ell,\epsilon)} (\bu)
= \arccos(\bu\bar\ell)+\frac{\epsilon}{\bar\ell}\left( \sqrt{1-\bu^2 \bar\ell^2}+\frac{1}{\sqrt{1-\bu^2 \bar\ell^2}}\right)\,.
\eeq
Finally $\ell$ is itself a function of $\epsilon$ when expressed as a function of the minimal orbit radius $R$ using Eq.~\eqref{ell-R} with $M$ replaced by $\epsilon M$. This further expansion of Eq.~\eqref{Phiepsilon} leads to
\beq
\Phi_{(\bar\ell,\epsilon)} (\bu)
= \arccos(\bu\bar R)+\frac{2}{\bar R}\frac{1-\frac{\bu\bar R}{2}-\frac{\bu^2\bar R^2}{2}}{ \sqrt{1-\bu^2 \bar R^2}}\epsilon +O(\epsilon^2)
\,,
\eeq
which coincides with Eq.~(2) of Ref.~\cite{Arifov:1969}.

As in the flat spacetime case, if two photons arrive at a circle of radius $r_0=M/\bu_0$ at two successive angles $\phi_1$ and $\phi_2$ with parameters $\bar\ell_1$ and $\bar\ell_2$ from the same point of emission $(r_*,\phi_*)$,  we can calculate the minimal distance angles $\phi^{(0)}(\bar\ell_1)$ and $\phi^{(0)}(\bar\ell_2)$, for example 
$
\phi_1-\phi^{(0)}(\bar\ell_1)
= \Phi_{(\bar\ell_1,\epsilon)} (\bu_0)
+O(\epsilon^2) \,,
$
and a similar relation for the second photon.
Then once we have $\phi_{\bar\ell_1}$ and $\phi_{\bar\ell_2}$ known in terms of $(r_0,\phi_1,\phi_2,\bar\ell_1,\bar\ell_2)$,
we have
\beq
\phi_*-\phi^{(0)}(\bar\ell_1)
= \Phi_{(\bar\ell_1,\epsilon)} (\bu_*)
+O(\epsilon^2)
\,,
\eeq
\beq
\phi_*-\phi^{(0)}(\bar\ell_2)
= \Phi_{(\bar\ell_2,\epsilon)} (\bu_*)
+O(\epsilon^2)
\,.
\eeq
Subtracting these two relations leads to
\begin{eqnarray}
\delta_{12}&=&
\phi^{(0)}(\bar\ell_2)-\phi^{(0)}(\bar\ell_1)
= 
\Phi_{(\bar\ell_1,\epsilon)} (\bu_*) - \Phi_{(\bar\ell_2,\epsilon)} (\bu_*)
+O(\epsilon^2)
\,,
\end{eqnarray}
which can be solved for $\bu_*$, leading to
\begin{eqnarray}
\bu_*
&=& 
\frac{|\sin \delta_{12}|}
{\sqrt{\bar\ell_1^2+\bar\ell_2^2-2\bar\ell_1\bar\ell_2 \cos  \delta_{12}}} 
+ \epsilon \frac{2(\bar\ell_1^2+\bar\ell_2^2) \cos \delta_{12}-\bar\ell_1\bar\ell_2 (1+3\cos^2 \delta_{12}) }
{\bar\ell_1\bar\ell_2 (\bar\ell_1^2+\bar\ell_2^2-2\bar\ell_1\bar\ell_2 \cos \delta_{12} )}\nonumber\\
&&
+O(\epsilon^2)
\,.
\end{eqnarray}
This result can be substituted in either of the previous two relations to determine $\phi_*$.
Higher-order approximation formulas can be obtained straightforwardly.

\end{document}